\documentclass{aa501}
\usepackage{epsf}

\begin{document}


\newcommand{\pF}{\mbox{$p_{\mbox{\raisebox{-0.3ex}{\scriptsize F}}}$}}  
\newcommand{\vph}[1]{\mbox{$\vphantom{#1}$}}  
\newcommand{\kB}{\mbox{$k_{\rm B}$}}           
\newcommand{\vF}{\mbox{$v_{\mbox{\raisebox{-0.3ex}{\scriptsize F}}}$}}  
\renewcommand{\arraystretch}{1.5}

\title{ Bulk viscosity in superfluid neutron star cores}
\subtitle{  III. Effects of $\Sigma^-$ hyperons }
\author{
        P.~Haensel\inst{1}
\and 
       K.P.~Levenfish\inst{2}\and D.G.~Yakovlev\inst{2}
        }
\institute{
 N.~Copernicus Astronomical Center,
       Bartycka 18, 00-716 Warszawa, Poland
\and
        Ioffe Physical Technical Institute, Politekhnicheskaya 26,
        194021 St.-Petersburg, Russia \\
        {\it e-mails: haensel@camk.edu.pl,
                      ksen@astro.ioffe.rssi.ru,
                      yak@astro.ioffe.rssi.ru} 
              }
\date{}
\offprints{P.Haensel: haensel@camk.edu.pl}

\titlerunning{ Bulk viscosity in superfluid neutron star cores}
\authorrunning{P.~Haensel et al.}

\abstract{
Bulk viscosity of neutron star cores containing hyperons
is studied taking into account  
non-equilibrium weak  process  $n+n \rightleftharpoons p+\Sigma^-$.
Rapid growth of the bulk viscosity
within the neutron star core associated with
switching on new reactions (modified Urca process, direct
Urca process, hyperon reactions) is analyzed.
The suppression of the bulk viscosity by superfluidity of
baryons is considered and found out to be very
important.
\keywords{ Stars: neutron --- dense matter}
}

\maketitle


\section{Introduction}
\label{sect-introduc}

The bulk viscosity of matter in the cores of neutron stars
has recently attracted great attention in connection with
damping of neutron star pulsations and gravitational
radiation driven instabilities, particularly ---
in 
damping of $r$-modes (e.g., Andersson \& Kokkotas \cite{ak00}). 
It is well known that the bulk viscosity is caused by
energy dissipation associated with weak-interaction 
non-equilibrium reactions in a pulsating dense matter. 
The reactions
and the bulk viscosity itself
depend sensitively on the composition 
of matter.

In the outermost part of the outer neutron star core
composed mainly of neutrons $n$ with admixture
of protons $p$, electrons $e$, and possibly muons $\mu$
bulk viscosity is mainly determined by the reactions
of non-equilibrium modified Urca process,
\begin{equation}
    n+N \to N+p+l+\bar{\nu}_l,\quad  p+N+l \to n+N+\nu_l,
\label{Murca}
\end{equation}
where $N$ stands for a nucleon ($n$ or $p$),
$l$ is an electron or a muon, and $\nu_l$ is 
an associated neutrino. The problem of damping neutron star
pulsations via modified Urca process in $npe$ matter
was analyzed long ago by Finzi \& Wolf (\cite{fw68})
(although the authors did not introduce the bulk viscosity
explicitly).
The bulk viscosity in $npe$ matter
was calculated by Sawyer (\cite{sawyer89}) and
in $npe\mu$ matter by Haensel et al.\ (\cite{hly01}, 
hereafter Paper II). 

Deeper in the core, at densities $\rho$ of a few $\rho_0$
($\rho_0=2.8 \times 10^{14}$ g cm$^{-3}$ is the
saturated nuclear matter density), direct Urca process
may be open (Lattimer et al.\ \cite{lpph91})
\begin{equation}
   n  \to p +l+\bar{\nu}_l, \quad p+l \to n+\nu_l.  
\label{Durca}
\end{equation}
It produces the bulk viscosity, which is typically 4--6
orders of magnitudes higher than that due to modified
Urca process. This bulk viscosity was studied by
Haensel \& Schaeffer (\cite{hs92}) for $npe$ matter
and by Haensel et al.\ (\cite{hly00}, hereafter Paper I)
for $npe\mu$ matter. Note that the idea 
of strong enhancement of vibrational dissipation in
the neutron stars 
via a weak process similar to direct Urca 
(beta decay and capture by quasinuceons in 
$npe$ matter containing pion condensate)
was put forward by Wang \& Lu (\cite{wl84}).
All the studies cited above
are focused on not too young neutron stars
which are fully transparent to neutrinos.
We will also restrict ourselves to this case.

At about the same densities, hyperons may appear in 
the neutron star cores (first of all, $\Sigma^-$ and
$\Lambda$ hyperons, and then $\Xi^0$, $\Xi^-$, $\Sigma^+$). 
To be specific, we will mainly consider $\Sigma^-$ and $\Lambda$
hyperons. Once appeared,
the hyperons may also initiate their own 
direct Urca processes (Prakash et al.\ \cite{pplp92})
giving additional contribution to
the bulk viscosity, nearly as high as
that due to nucleon direct Urca process (\ref{Durca}).
However, direct non-leptonic hyperon collisions 
which go via weak interaction
(with strangeness non-conservation) such as
\begin{eqnarray} 
   n+n \rightleftharpoons p+\Sigma^-, && 
\label{Sigma}\\
   \Lambda + N  \rightleftharpoons n + N, &&
\label{Lambda}
\end{eqnarray}
are much more efficient. They may increase
the bulk viscosity  by several orders
of magnitude above the direct-Urca level. 
The effect was analyzed by Langer \&
Cameron (\cite{lc69}) and Jones (\cite{jones71,jones01a,jones01b}).
Analogous effect in quark matter was studied by 
several authors. The enhancement of
the vibrational dissipation  via
non-lepton strangeness-changing quark collisions
was considered by Wang \& Lu (\cite{wl84}). The appropriate
bulk viscosity was calculated by Sawyer (\cite{sawyer89b}),
Madsen (\cite{madsen92}), Goyal et al.\ (\cite{goyaletal94}),
and Dai \& Lu (\cite{dl96}). 

Calculation of the bulk viscosity limited by non-leptonic
processes in hyperon matter is a complicated problem.
There are a number of processes of comparable
efficiency. The matrix elements can easily be calculated
in the approximation of bare particles and exact SU(3) symmetry, 
and appear
to be nonzero for some processes, e.g., (\ref{Sigma}),
but are zero for the others, e.g.,
(\ref{Lambda}). However,
experimental data on 
the lifetime of $\Lambda$ in massive
hypernuclei indicate (e.g., Jones \cite{jones01b}
and references therein) that process (\ref{Lambda})
(with $N=n$) is nearly as efficient as ``bare-particle''
process (\ref{Sigma}). Calculation of the matrix elements
for ``dressed'' particles is complicated and model
dependent; additional complications arise  --- even 
in the in-vacuum case  --- due to the 
SU(3) symmetry breaking (Savage \& Walden \cite{sw97}). 

Another complication is introduced by superfluidity
of neutron-star matter. It is well known that
neutrons, protons and other baryons  
may be superfluid due to attractive part of
strong baryon-baryon interaction. Superfluidity
of neutrons and protons has been studied in numerous
papers (as reviewed, for instance, by
Yakovlev et al.\ \cite{yls99} and Lombardo \& Schulze \cite{ls01}).
Hyperons can also be in superfluid state
as discussed, e.g., by Balberg \& Barnea (\cite{bb98}).
Critical temperatures $T_c$ of baryon superfluidities
are very sensitive to the model of strong interaction
and to many-body theory employed in microscopic calculations.
Their typical values range from $10^8$ to $10^{10}$ K.
They are density dependent, and they mainly decrease
with $\rho$ at densities higher than several $\rho_0$. 

The effects of superfluidity of nucleons on the bulk viscosity
associated with direct and modified Urca processes in $npe\mu$
matter were considered in Papers I and II.
It was shown that the superfluidity may drastically
reduce the bulk viscosity and, hence, the damping of
neutron star pulsations.

In this paper we propose a simple solvable model
of the bulk viscosity in hyperonic matter (Sect.\ 2) due
to process (\ref{Sigma}) and study (Sect.\ 3) the effects
of possible superfluidity of $n$, $p$, and $\Sigma^-$
on this bulk viscosity. In Sect.\ 4 we discuss
density and temperature dependence of the bulk viscosity in
non-superfluid and superfluid neutron star cores.

\section{Bulk viscosity of non-superfluid matter}
\label{sect-nonsup}

\subsection{Model}

Consider non-superfluid hyperonic stellar matter in
the core of a neutron star pulsating with
a typical frequency $\omega \sim 10^3-10^4$ s$^{-1}$.
In the presence of hyperons the contribution of direct Urca 
and modified Urca processes, (\ref{Durca}) and (\ref{Murca}),
into the bulk viscosity may be neglected. It is sufficient
to include the non-leptonic weak-interaction processes
(\ref{Sigma}) and (\ref{Lambda}). For the sake of
simplicity, let us take into account process
(\ref{Sigma}) alone although we assume that matter
may contain not only $\Sigma^-$ but other hyperons.
The advantage of this model is that it can be solved
analytically. We will compare it with other models
in Sect.\ 2.4. 

\subsection{ Matrix element of $n+n \to p + \Sigma^-$}

Let us start with the matrix element $M$ in the
``bare-particle'' approximation. The process
is described by two diagrams with the states of two neutrons
interchanged. Accordingly,
$M=M^{\rm(I)}+M^{\rm (II)}$, and  ($\hbar= c = k_{\rm B} =1$)
\begin{eqnarray}
M^{\rm (I)}&=&{\cal A}\left[
    \bar{u}_p \gamma_\lambda
    (1+ C \gamma_5) u_n \right]
    \left[
     \bar{u}_{\Sigma} \gamma^\lambda
    (1+ C' \gamma_5) u_{n^\prime} \right] ,
\nonumber \\
M^{\rm (II)} &=& - {\cal A} \left[
    \bar{u}_p \gamma_\lambda
    (1+ C \gamma_5) u_{n^{\prime}} \right]
    \left[ \bar{u}_\Sigma \gamma^\lambda
    (1+ C' \gamma_5) u_n \right].
\label{M2} 
\end{eqnarray}
In this case $u_i$ is a standard bispinor, $\bar{u}_i$ is 
its Dirac conjugate
($i=n,\,n^\prime,\, p,\, \Sigma$; $\bar{u}_i u_i=2 m_i$,
where $m_i$ is a bare-particle mass),
$\gamma^\lambda$ is a Dirac's gamma-matrix, and 
\begin{equation}
   {\cal A} = -\, \frac{G_{\rm F}}{\sqrt{2}} \,
       \sin \theta_{\rm C} \, \cos \theta_{\rm C}\, .
\label{amplitude}
\end{equation}
Furthermore, $G_{\rm F}=1.436\times 10^{-49}$ erg cm$^3$ is the Fermi 
weak coupling constant; $\theta_{\rm C}$ is the Cabibbo
angle ($\sin \theta_{\rm C}=0.231$);
$C=F+D$, $C'=F-D$,
where $D \approx 0.756$ and $F \approx 0.477$
are the reduced symmetric and antisymmetric coupling
constants (e.g., Prakash et al.\ \cite{pplp92}).

Using the standard technique in the limit of
non-relativistic baryons we sum $|M|^2$ over particle
spin states and obtain
\begin{equation}
   \sum_{\rm spins} |M|^2 =
   64\, |{\cal A}|^2 \, \chi \: m^2_n\,  
   m_p \, m_\Sigma , \quad \chi =(1+3\,C C')^2.
\label{matrix} 
\end{equation}
This expression coincides with that which can
be deduced from the recent results of Jones (\cite{jones01b}).
In the previous papers Jones (\cite{jones71,jones01a}) reported
analogous expression but with $\chi' \! =\!(1\!-\!3\,C C')^2$ instead of $\chi$. 
Numerically, replacing minus with plus makes a great
difference due to almost total compensation of the terms
in $\chi$: $\chi \approx 0.001$ and $\chi' \approx 4.13$.
Because of the strong compensation
we cannot rely on the bare-particle approximation.
Let us assume that a more evolved calculation based on
dressed-particle technique will lead to the same Eq.\ (\ref{matrix})
but with the value of $\chi$ renormalized by medium effects.
Accordingly we will treat $\chi$ as a free parameter and, 
to be specific, we will set $\chi=0.1$.

\subsection{Non-equilibrium  rate}

Due to very frequent interparticle collisions,
dense stellar matter almost instantaneously (on microscopic time scales)
achieves a quasiequilibrium state with certain temperature $T$ and 
chemical potentials $\mu_i$ of various particle species $i$.
Relaxation to the full thermodynamic
(``chemical'') equilibrium lasts much longer
since it realizes through much slower weak interaction processes.

In the case of process (\ref{Sigma}) the chemical equilibrium implies
$2\mu_n=\mu_p + \mu_\Sigma$. In the chemical equilibrium
the rates $[\,\rm{cm}^{-3}$ s$^{-1}]$ of the direct and 
inverse reactions of the process are balanced,
$\Gamma=\bar{\Gamma}$.
In a pulsating star, the chemical equilibrium is
violated  ($\Gamma \neq \bar{\Gamma}$)
which can be described by the lag
of {\it instantaneous} chemical potentials,
\begin{equation}
    \eta= 2\mu_n-\mu_p-\mu_\Sigma.
\label{eta}
\end{equation}
We adopt the standard
assumption (e.g., Sawyer \cite{sawyer89}) that deviations
from the chemical equilibrium are small, 
$|\eta | \ll T$.  If so we can use
the {\it  linear approximation}
\begin{equation}
     \Delta \Gamma \equiv\Gamma-\bar{\Gamma} = -\lambda \eta,
\label{lambda_def}
\end{equation}
where $\lambda$ determines the
bulk viscosity (Sect.\ 2.4). Our definition
of $\lambda$ is the same as in Sawyer (\cite{sawyer89}). 
Thus defined, $\lambda$ is negative.

Let us calculate the rate $\Gamma$ 
of the direct reaction, $nn \to p\Sigma^-$, of the process.
In the non-relativistic approximation 
we have  ($\hbar = c = k_{\rm B} = 1$):
\begin{eqnarray}
\Gamma &  = & \int 
           \frac{{\rm d} \vec{p}_n}{2 m_n (2\pi)^3 }\,
           \frac{{\rm d} \vec{p}'_n}{2 m_n (2\pi)^3 }\,    
           \frac{{\rm d} \vec{p}_p}{2 m_p (2\pi)^3 } \,
           \frac{{\rm d} \vec{p}_\Sigma}{2 m_\Sigma (2\pi)^3}
\nonumber \\
       &   & \times \, \frac{1}{2} \sum_{\rm spins} |M|^2  \,             
            f_n \,f'_n\,  (1 -f_p) (1 - f_\Sigma) \, (2 \pi)^4  
\nonumber \\
       &   & \times \, 
           \delta(\varepsilon_n + \varepsilon'_n 
           - \varepsilon_p - \varepsilon_\Sigma)\,   
           \delta(\vec{p}_n+\vec{p}'_n-\vec{p}_p-\vec{p}_\Sigma),
\label{Gamma}
\end{eqnarray}
where $\vec{p}_i$ is the particle momentum and
$\varepsilon_i$ is its energy.
The symmetry factor 
${1 \over 2}$ before summation sign 
excludes double counting of the same collisions of identical 
neutrons; $f_i= \{1+\exp[(\varepsilon_i-\mu_i)/T]\}^{-1}$ 
is the Fermi-Dirac function.

Evaluation of $\Gamma$ is standard (e.g.\ Shapiro \& Teukolsky \cite{st83})
and takes advantage of strong degeneracy of reacting particles
in neutron star matter. 
The multidimensional integral
is decomposed into the energy and 
angular integrals. All momenta $\vec{p}_i$ 
are placed on the appropriate Fermi spheres 
wherever possible. 
Introducing the dimensionless quantities
\begin{equation}
   x_i=\frac{\varepsilon_i - \mu_i}{T}, \quad  \xi = {\eta \over T},
\label{dimensionless}
\end{equation}
we can rewrite the reaction rate as  $\Gamma = \Gamma^{(0)} I$,
with
\begin{equation}
    I = \left[ \, \prod^4_{i=1}   \int_{-\infty}^{+\infty}  
        {\rm d}x_i \: f(x_i) \right]\, 
         \delta \left( \sum^4_{i=1} x_i +\xi \right),
\label{Idef}
\end{equation}
where the blocking factors  $(1-f(x))$ 
are transformed into the Fermi-Dirac 
functions $f(x)$ by replacing integration variables $x\to -x$, 
and the typical reaction rate $\Gamma^{(0)}$ is
defined as (in ordinary physical units)
\begin{eqnarray}
   \Gamma^{(0)}  &= &  \frac{4}{(2\pi)^5\, \hbar^{10}} \; G_{\rm F}^2\: 
           \sin^2 \theta_{\rm C}\, \cos^2 \theta_{\rm C} \, \chi
\nonumber \\
          & & \times \;  m^{\ast 2}_n\, m^\ast_p\, m_\Sigma^\ast \;
           p_{{\rm F}\Sigma}\, k_{\rm B}^3 \,T^3
\nonumber \\
          & & \approx \; 2.15 \times 10^{38}
          \,  \left( {m^{\ast}_n  \over m_n} \right)^{\!2}
\nonumber \\
      & &  \times
             \;
           {m^\ast_p \, m^\ast_\Sigma  \over m_p \, m_\Sigma}
            \left( {n_\Sigma \over 1\,{\rm fm}^{-3}} \right)^{\!1/3} 
	    T_9^3 \, \chi \;\,
           {\rm cm}^{-3}~{\rm s}^{-1}.
\label{Gamma0}   
\end{eqnarray}
In this case $m_i^\ast$ is an effective baryon mass
in dense matter,
$p_{{\rm F}\Sigma}$ is the Fermi
momentum of $\Sigma^-$ hyperons,
$n_\Sigma$ is their number density,
$T_9=T/(10^9\, {\rm K})$. Note that in Eq.\ (\ref{Gamma0})
we have used the angular integral calculated
under the assumptions $p_{{\rm F}\Sigma} + p_{{\rm F}p}
< 2 p_{{\rm F}n}$ and $p_{{\rm F}\Sigma}< p_{{\rm F}p}$
which are usually fulfilled in hyperonic matter
($p_{{\rm F}i}$ being Fermi momentum of particle species $i$).

The integral  $I$, Eq.\ (\ref{Idef}), is:  
\begin{equation}
   I= {{\rm e}^\xi \over {\rm e}^\xi -1 } \;
   {\xi \over 6} \;\left(4\pi^2 + \xi^2 \right).
\label{I}
\end{equation}

The rate $\bar{\Gamma} = \Gamma^{(0)}\bar{ I}$ of the inverse reaction,
$\Sigma^-p \to nn$,
is obtained from $\Gamma$, Eqs. (\ref{Gamma}) and (\ref{Idef}),
by replacing $\xi \to -\xi$.
Then for $|\xi| \ll 1$
\begin{equation}
       \Delta \Gamma  =  \Gamma^{(0)}\, \Delta I,\quad
       \Delta I = {2 \pi^2 \over 3} \, \xi.
\label{DeltaGamma}
\end{equation}
Finally, from
Eqs.\ (\ref{lambda_def}) and  (\ref{DeltaGamma}) we obtain
\begin{equation}
     |\lambda|={\Gamma^{(0)}\over k_{\rm B} T}\;{\Delta I \over \xi} .
\label{lambda}
\end{equation}
In non-superfluid matter
$ |\lambda_0 | =2\pi^2\Gamma^{(0)}/(3 k_{\rm B}T)$.

\subsection{Bulk viscosity}

The bulk viscosity $\zeta_\Sigma$ due to
the hyperon process (\ref{Sigma}) is calculated in
analogy with that due to the modified
or direct Urca process (Sawyer \cite{sawyer89};
Haensel \& Schaeffer \cite{hs92}). The result is
\begin{equation}
    \zeta_\Sigma= {C^2 n_b^2 \over |\lambda| B^2} \:
    {1 \over 1+a^2}, \quad a \equiv {\omega n_b \over |\lambda| B},
\label{hvis}
\end{equation}
where $n_b$ is the number density of baryons, and
\begin{equation}
    B= { \partial \eta \over \partial X_\Sigma}, \quad
    C= n_b {\partial \eta \over \partial n_b} =
       - { 1 \over n_b} \, {\partial P \over \partial X_\Sigma}.
\label{BC}
\end{equation}
In this case $P$ is the pressure and $X_\Sigma = n_\Sigma / n_b$
is the fraction of $\Sigma^-$ hyperons. The quantities $B$ and
$C$ can be calculated numerically for a given equation of
state.

The bulk viscosity depends on the frequency $\omega$ 
of neutron star pulsations. 
Using the results of Sect.\ 2.3 the dynamical parameter $a$ can be
written as
\begin{eqnarray}
     a & \approx & 6.09 \left(\! {m_n \over m_n^\ast}\! \right)^{\!2}
        {m_p \over m_p^\ast}\, {m_\Sigma \over m_\Sigma^\ast}         \,
        {\omega_4 \, \over  T_9^2 \,\chi} 
\nonumber \\ 
         & & \times \; \left( 100~{\rm MeV} \over B \right)  
           \left( {n_b \over 1\, {\rm fm}^{-3}}  \right)
	\left({1 \, {\rm fm}^{-3} \over n_\Sigma} \right)^{1/3},
\label{a}
\end{eqnarray}
where 
$\omega_4 = \omega /(10^4~{\rm s}^{-1})$.
For typical values $T \sim 10^8-10^9$ K,
$\omega_4 \sim 1$,
$n_b \sim 1$ fm$^{-3}$,
$n_\Sigma \ll n_b$, $m_i^\ast \sim 0.7 m_i$,
$B \sim 100$ MeV, $\chi \sim 0.1$
we have $a \gg 1$. Then we may use the {\it high-frequency limit}
in which $\zeta_\Sigma$ is independent of $B$ and inversely 
proportional to $\omega^2$: 
\begin{eqnarray}
   \zeta_\Sigma &=&
   {2\over3\,(2\pi)^3}\,
       { G_{\rm F}^2 \,
        m^{\ast 2}_n\; m^\ast_p\; m^\ast_\Sigma\; C^2 \chi\; p_{{\rm F}\Sigma}
           \over \hbar^{10} \omega^2 }
\nonumber \\
   & & \times \sin^2 \!\theta_{\rm C}\,\cos^2 \theta_{\rm C}\: (k_{\rm B}T)^2
\nonumber \\
    & \approx & 2.63  \times  10^{30} \: T_9^2 \, \omega^{-2}_4
       \, \chi \, \left( {n_\Sigma \over 1\, \rm{fm}^{-3}} \right)^{1/3}
\nonumber \\
    && \times  \left( {m^{\ast}_n \over  m_n} \right)^{\!2}
       {m^\ast_p \over  m_p}\:
    {m^\ast_\Sigma \over  m_\Sigma}
   \left( {C \over 100\,{\rm MeV}} \!\right)^{\!2} \;\:
 {{\rm g~cm}^{-1}~{\rm s}^{-1}}.
\label{zeta0}
\end{eqnarray}
If, due to interplay of parameters, $a \la 1$ one can use
more general Eq.\ (\ref{hvis}). For instance, we would
have $a \la 1$ for the same parameters as above
but at higher temperatures, $T \ga 10^{10}$ K
(Sect.\ 4).
We could have $a \la 1$ even below $\sim 10^{10}$ K
if the phenomenological constant $\chi$ is higher than
the adopted value $\chi=0.1$.

In the absence of hyperons the bulk viscosity is determined
by direct or modified Urca processes (Sect.\ 1).
These processes are much slower than hyperonic ones.
They can certainly be described in the high-frequency
approximation in which partial bulk viscosities due to
various processes are summed together into the total
bulk viscosity (e.g., Papers I and II). Thus we will add
contributions from direct and modified Urca processes
whenever necessary in our numerical examples in Sect.\ 4.

Note that all the studies of bulk
viscosity of hyperonic matter performed
so far are approximate.
The subject was introduced by Langer \&
Cameron (\cite{lc69}) who estimated
dumping of neutron star vibrations but did not calculate
the bulk viscosity itself. Jones (\cite{jones71,jones01a})
calculated effective $\Sigma^-$ hyperon relaxation
times and estimated the bulk viscosity but did not
evaluate it exactly for any selected model of dense
matter. Recently Jones (\cite{jones01b}) analyzed
the bulk viscosity of hyperonic matter taking into account
a number of hyperonic processes
but also restricted himself to the order-of-magnitude
estimates.

Our approach is also simplified since we take
into account the only one hyperonic process (\ref{Sigma}) 
and neglect the others. Even in this case we are forced
to introduce the phenomenological parameter $\chi$ (Sect.\ 2.2)
to describe the reaction rate. The advantage of our
model is that, once this parameter is specified,
we can easily calculate the bulk viscosity (as illustrated
in Sect.\ 4) and introduce the effects of
superfluidity (Sect.\ 3, 4). Technically, it would be easy 
to incorporate the contribution
of process (\ref{Lambda}) as well
as of other hyperonic processes (Sect.\ 1). However, for any
new process we need
its own phenomenological parameter
(similar to $\chi$) which is currently unknown. 
Generally,
in the presence of several hyperonic processes, the bulk viscosity
cannot be described by a simple analytical expression
analogous to Eq.\ (\ref{hvis}). 
Nevertheless, in the high-frequency limit the contributions
from different processes are additive and
it will be sufficient to add new contributions to that
given by Eq.\ (\ref{zeta0}).
Thus we prefer to use
our simplified model rather than extend it introducing large
uncertainties.

\section{Bulk viscosity of superfluid matter}

\subsection{Baryon pairing in dense matter} 

Now consider the effects of baryon superfluidity
on the bulk viscosity associated with process (\ref{Sigma}).
According to microscopic theories 
(reviewed, e.g., by
Yakovlev et al.\ \cite{yls99} and Lombardo \& Schulze \cite{ls01})
at supranuclear densities
(at which hyperons appear in dense matter)
neutrons may undergo triplet-state ($^3$P$_2$)
Cooper pairing while protons
may undergo singlet-state ($^1$S$_0$) pairing. 
As discussed in Sect.\ 1 
microscopic calculations of
the nucleon gaps (critical temperatures) are very model
dependent. Current knowledge of hyperon interaction
in dense matter is poor and therefore microscopic
theory of hyperon pairing is even much more uncertain.
Since the number density
of hyperons is typically not too large
it is possible to expect that such a pairing,
if available, is produced by singlet-state hyperon
interaction. Some authors (e.g., Balberg \& Barnea \cite{bb98})
calculated singlet-state gaps for $\Lambda$ hyperons.
We assume also singlet-state pairing
of $\Sigma^-$ hyperons and consider the bulk viscosity
of matter in which $n$, $p$ and $\Sigma^-$ may form
three superfluids. Since the critical temperatures
$T_{cn}$, $T_{cp}$ and $T_{c\Sigma}$ are uncertain
we will treat these temperatures as
arbitrary parameters.

Microscopically, superfluidity introduces a gap
$\delta$ into momentum dependence of the baryon energy,
$\varepsilon(\vec{p})$. Near the Fermi surface
($|p - p_{\rm F}| \ll p_{\rm F}$) we have
\begin{eqnarray}
\varepsilon & = & \mu - \sqrt{\delta^2+v^2_{\rm F}\, (p-p_{\rm F})^2}
                                \quad  {\rm at }\;\; p<p_{\rm F},
\nonumber\\
\varepsilon & = & \mu + \sqrt{\delta^2+v^2_{\rm F} (p-p_{\rm F})^2}
                                \quad {\rm at }\;\; p\ge p_{\rm F},
\label{dispersion}
\end{eqnarray}
where $v_{\rm F}$ is the Fermi velocity.
The gap $\delta$ is isotropic (independent
of orientation of $\vec{p}$ with respect to the 
spin quantization axis) for
singlet-state pairing but anisotropic for triplet-state
pairing. Strict calculation of the bulk viscosity
with anisotropic gap is complicated. We will
adopt an approximate treatment
of triplet-state pairing 
(with zero projection of total angular momentum of Cooper pairs
onto the spin quantization axis)
proposed by Baiko et al.\ (\cite{bhy01})
for calculating diffusive thermal conductivity of neutrons.
In this approximation the gap is artificially considered as isotropic
in microscopic calculations 
but in the final expressions it
is related to temperature in the same
way as the minimum value of the anisotropic gap
on the Fermi surface. 

It is convenient to introduce the dimensionless quantities
\begin{equation}
   \tau={T\over T_c},
     \quad
   y =  {\delta(T) \over T}, 
     \quad 
   z = {\rm sign}(x)\,\sqrt{ x^2 + y^2}.
\label{dimless}
\end{equation}
For the singlet-state pairing (case A in notations of 
Yakovlev et al.\ \cite{yls99})
the dependence of 
$y$ 
on $\tau$ can be fitted as
\begin{equation}
   y_{\rm A}=\sqrt{1-\tau }\,
   \left( 1.456 - { 0.157 \over \sqrt{\tau}} + {1.764 \over \tau}  
   \right) ,
\label{yA}
\end{equation}
while for the triplet-state pairing (case B)
\begin{equation}
   y_{\rm B}=\sqrt{1-\tau }\,
   \left( 0.7893  + {1.188 \over \tau}  
   \right)  .
\label{yB}
\end{equation}

\subsection{Superfluid reduction factors} 

We consider the effects of superfluidity on the bulk viscosity
in the same manner as in Papers I and II and omit technical
details described in these papers. 
Following Papers I and II we assume that all constituents of matter
participate in stellar pulsations with the same macroscopic 
velocity (as in the first-sound waves). Then
the damping of pulsations is described by one coefficient of bulk viscosity
$\zeta$. The effects of superfluidity are included
by introducing superfluid gaps into the reaction rates,
$\Gamma$ and $\bar{\Gamma}$, Eq.\ (\ref{Gamma}),  
through the dispersion relations, Eq.\ (\ref{dispersion}).
These effects influence mainly the only parameter $\lambda$
in Eq.\ (\ref{hvis}). Quite generally, we can write
\begin{equation}
   \lambda = \lambda_0 \, R,
\label{lambda_SF}
\end{equation}
where $\lambda_0$ refers to non-superfluid matter,
Eq.\ (\ref{lambda}), and $R$ is a factor 
which describes the superfluid effects.
The latter factor depends  on the three parameters,
$R=R(y_n,\, y_p, \, y_{\Sigma})$,
which are the dimensionless gaps of neutrons, protons, and 
$\Sigma^-$ hyperons. Obviously, $R\!=\!1$
if all these baryons are normal ($y_n\!=\!y_p\!=\!y_\Sigma\!=\!0$).
Calculations show that one always has $R<1$ in the presence
of at least one superfluidity.

Using Eqs.\ (\ref{hvis}) and (\ref{lambda_SF})
we can write the hyperon bulk viscosity in 
superfluid matter in the form
\begin{equation}
    \zeta_\Sigma= {C^2 n_b^2 \over |\lambda_0|  B^2 R} \:
    {1 \over 1+a^2}, \quad a = {\omega n_b \over |\lambda| B}
    ={a_0 \over R},
\label{hvis_SF}
\end{equation}
where $a_0$ is the non-superfluid value of $a$ 
given by Eq.\ (\ref{a}).

In the high-frequency limit (Sect.\ 2.4),
which is often realized in neutron star matter,
we have
$\zeta_\Sigma \propto \lambda$, i.e.,
\begin{equation}
  \zeta_\Sigma = \zeta_0 \, R,
\label{zeta_SF}
\end{equation}
where $\zeta_0$ is the bulk viscosity of non-superfluid
matter, Eq.\ (\ref{zeta0}). Accordingly, superfluidity
{\it suppresses} the high-frequency bulk viscosity.
On the contrary, it {\it enhances}
the static ($\omega=0$) bulk
viscosity $\zeta_\Sigma \propto 1/\lambda \propto 1/R$.
Moreover, superfluidity increases the dynamical factor
$a$ and widens thus the range of plasma parameters
where the bulk viscosity operates in the high-frequency regime.

Under our assumptions superfluidity modifies only the integral $\Delta I$
in the factor $\lambda$ given by Eq.\ (\ref{lambda}). 
To generalize $\Delta I$ to the superfluid case it is sufficient to
replace $x_i \to z_i$  
in the all functions
under the integral in Eq.\ (\ref{Idef}).
Then $R$ can be written as 
\begin{equation}
   R={\Delta I \over \Delta I_0} =
   {3 \over \pi^2} {\partial \over \partial \xi}
   \left[ \prod^4_{i=1}   \int  {\rm d} x_i \, f(z_i) \right]
       \delta \left( \sum^4_{i=1} z_i + \xi  \right) 
\label{RR}
\end{equation}
in the limit of $\xi \to 0$. Here $\Delta I_0$ is 
the value of $\Delta I$ calculated for normal matter, Eq.\ (\ref{DeltaGamma}).

We have composed a code which calculates $R$ numerically
in the presence of all three superfluids. The results 
will be presented in Sect.\ 4. 
Here we mention some limiting cases in which
evaluation of $R$ is simplified.

\subsection{Superfluidity of protons or $\Sigma^-$ hyperons}
%
The cases in which either protons or $\Sigma^-$ hyperons are superfluid
are similar. Let, for example, neutrons and $\Sigma^-$
be normal while protons
undergo $^1$S$_0$ Cooper pairing.
Accordingly, $R= R_p$ 
depends on the only parameter $y = y_{{\rm A}p}$.
For a strong superfluidity ($\tau = T/T_{cp}  \ll  1$, 
$y  \gg  1$)
the asymptote is
\begin{equation}
  R_p={3\over \pi^2} \, \sqrt{\pi y \over 2} \;
  \left( 
{y^2\over 2}  
 + {y\over 2} + {\pi^2\over 6} 
  \right)
\:{\rm e}^{-y}.
\label{Rs-asy}
\end{equation}
We have calculated $R_p$ in a wide range of $y$ and
proposed  the fit to the numerical data (with
the maximum error $\la 0.5$\%) which reproduces also 
the leading term of the asymptote, Eq.\ 
(\ref{Rs-asy}):
\begin{equation}
 R_p  = {a^{5/4} + b^{1/2} \over 2} \: 
      \exp \left( 0.5068-\sqrt{0.5068^2  + y^2} \right)  , 
\label{Rs-fit}
\end{equation}
where $a=1+0.3118\, y^2$ and $b=1+2.556\, y^2$.
If $\Sigma^-$ hyperons are superfluid instead of protons, the
expressions for $R$ are the same but $y=y_{{\rm A}\Sigma}$.

\subsection{Superfluidity of protons and $\Sigma^-$ hyperons}

If neutrons are normal but protons and $\Sigma^-$ hyperons are 
superfluid  $R = R_{p\Sigma}(y_p,y_\Sigma)$ depends on
$y_p=y_{{\rm A}p}$ and $y_\Sigma=y_{{\rm A}\Sigma}$. We have determined
the asymptote of $R_{p\Sigma}$ at large $y_p$ and $y_\Sigma$.
Let $Y$ be the larger gap, 
 $Y\! = \! \max \{y_\Sigma, y_p\}$, and
$y_0 \!= \! \min\{ y_\Sigma, y_p\}$.
At  $Y \! -y_0 \! \gg \! \sqrt{Y} \! \gg \! 1$ the asymptote reads
\begin{eqnarray}
 R_{p\Sigma}& =&{3\over \pi^2} \, \sqrt{\pi Y\over 2} \; \, {\rm e}^{-Y}
                \left[
                {Y+1 \over 2} \;\sqrt{Y^2 - y_0^2}  \right.
\nonumber \\
   && - \left.
          {y_0^2 \over 2} \, \ln \left( {Y+\sqrt{Y^2-y_0^2}\over y_0}\right)
        \right] .
\label{Rps-asy}
\end{eqnarray}
If $y_0 \to 0$ then Eq.\ (\ref{Rps-asy}) reproduces 
the leading term of the asymptote (\ref{Rs-asy}).
To prove this
one should consider Eq.\ (\ref{Rps-asy}) at
$1 \ll \sqrt{Y} \ll (Y - y_0) \ll Y$
and expand the logarithm in Eq.\ (\ref{Rps-asy})
in powers of $\sqrt{Y^2 -y_0^2}/Y \ll 1$. 
 
Equation (\ref{Rps-asy}) becomes invalid at $y_0 \to Y$.
In this case $R_{p\Sigma}(y_p,y_\Sigma) \approx R_n(Y)$,
where $R_n$ is described below.

\subsection{Superfluidity of neutrons}

Now let neutrons be superfluid while protons and $\Sigma^-$ hyperons
not. For a strong superfluidity ($\tau=T/T_{cn} \ll 1$, 
$y=y_{{\rm B}n} \gg 1$)
we get
\begin{equation}
  R_n = {6\,\gamma \,y \over \pi^2}\, {\rm e}^{-y}, \;\;\;
      \gamma= 
     \int_{\raisebox{-6pt}{$\!\!\!\!\scriptstyle 0$}}^{
              \raisebox{4pt}{$\!\!\!\scriptstyle \infty$}}  \!\!\!
    \int_{\raisebox{-6pt}{$\!\!\!\!\scriptstyle 0$}}^{
              \raisebox{4pt}{$\!\!\!\scriptstyle \infty$}}  
             \!
           { {\rm d} q  \,{\rm d} q^{\prime} \,  (q^{\prime 2}-q^2)  \over 
          {\rm e\,}^{ q^{\prime 2}} - {\rm e\,}^{ q^2 } }
          = 1.413.
\label{Rn-asy}
\end{equation}
We have calculated $R_n$ numerically in a wide range of
$y$ and proposed the fit (with the maximum error $\sim 0.2$\%):
\begin{eqnarray}
  R_n &=& \left(0.6192 + \sqrt{0.3808^2 + 0.1561\, y^2}\right)
\nonumber \\
      & & \times \; \exp \left( 0.7756 - \sqrt{0.7756^2+y^2} \right)
         + 0.18766 \; y^2 
\nonumber \\
      & & \times \;\exp \left(1.7755 -\sqrt{1.7755^2 +4y^2} \right).
\label{Rn-fit}
\end{eqnarray}
%

\section{Results and discussion}

\subsection{Non-superfluid matter} 
 
For illustration, we use the equation of state of matter
in the neutron star core, proposed by Glendenning 
(\cite{glendenning85}) in the frame of relativistic mean field theory.
Specifically, we adopt case 3 considered by Glendenning
in which the appearance of $n$, $p$, $e$, $\mu$, $\Sigma^-$,
and $\Lambda$ is allowed.
(Note a misprint: numerical values of the parameters $b$ and $c$ of the
Glendenning (1985) model should be replaced as $b \to b/3$
and $c \to c/4$).  
In this model, muons appear at the baryon number density
$n_b=0.110$ fm$^{-3}$ (at $\rho=1.86 \times 10^{14}$ g cm$^{-3}$);
$\Lambda$ hyperons appear at $n_b=0.310$ fm$^{-3}$
($\rho=5.51\times 10^{14}$ g cm$^{-3}$); and
$\Sigma^-$ hyperons appear at $n_b=0.319$ fm$^{-3}$
($\rho=5.69\times 10^{14}$ g cm$^{-3}$). 
The density dependence of the fractions of various
particles is shown in Fig.\ 9 of Glendenning (\cite{glendenning85}).
Let us remind that saturation density of nuclear
matter $\rho_0 \approx 2.8 \times 10^{14}$ g cm$^{-3}$
corresponds to $n_{b0} \approx 0.16$ fm$^{-3}$.

\begin{figure}
\centering
\epsfysize=86mm
\epsffile[85 290 470 660 ]{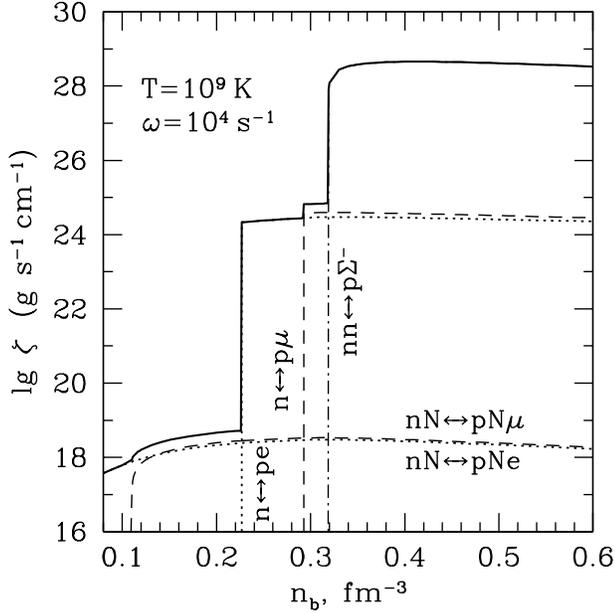}
\caption{
Density dependence of partial bulk viscosities
associated with various processes (indicated near the curves)
at $T=10^9$ K and $\omega=10^4$ s$^{-1}$
in non-superfluid matter. Dotted and dashed lines refer 
to Urca processes
involving electrons and muons, respectively; 
dot-and-dashed line refers to
hyperon process (\ref{Sigma}).
Thick solid line is the total bulk viscosity.
}
\label{fig1}
\end{figure}

Figure 1 shows the partial bulk viscosities and the total bulk viscosity
versus $n_b$ at $T=10^9$ K for stellar vibration frequency
$\omega=10^4$ s$^{-1}$. One can see three density intervals
where the bulk viscosity is drastically different.

At {\it low densities}, $n_b\! <\! 0.227$ fm$^{-3}$, the bulk viscosity
is determined by modified Urca processes (Paper II). For
$n_b\! <\! 0.110$ fm$^{-3}$ it is produced by neutron and proton
branches of Urca process involving electrons
(processes (\ref{Murca}) with $N=n$ or $p$ and with $l=e$).
At higher $n_b$ muons are created and muonic modified Urca processes
(\ref{Murca}) (again with $N\!=\!n$ or $p$ but now with $l\!=\!\mu$)
introduce comparable contribution. Note that Eq.\ (29)
of Paper II for the angular integral $A_{pl0}$ of the
proton branch of modified Urca process is actually valid at not
too high densities, as long as
$p_{{\rm F}n}\!>\! 3 p_{{\rm F}p} \!-\! p_{{\rm F}l}$. For higher
densities, it is replaced~with
\begin{equation}
    A_{pl0}= {(4 \pi)^5 \over 4 p_{{\rm F}p}^2 } \,
    \left( {3 \over p_{{\rm F}n} } - {1 \over p_{{\rm F}p} } \right),
\label{Apl0}
\end{equation}
which was neglected in Paper II (this replacement
has no noticeable effect on the values of bulk viscosity).

At {\it intermediate densities} 
($0.227$ fm$^{-3} < n_b < 0.319$ fm$^{-3}$)
the main contribution into the bulk viscosity comes from
direct Urca processes (Paper I). 
As long as $n_b < 0.293$ fm$^{-3}$ the only one direct
Urca process (\ref{Durca}) operates with $l=e$ while
at higher $n_b$ the other one with $l=\mu$ is switched on;
it makes comparable contribution.
We see that direct Urca processes 
at intermediate densities amplify the
bulk viscosity by more than five orders of magnitude
as compared to the low-density case.

Finally, at {\it high densities} ($n_b > 0.319$ fm$^{-3}$),
according to the results of Sect.\ 2,
the bulk viscosity increases further by about four 
orders of magnitude under the action of non-leptonic 
process (\ref{Sigma}) involving $\Sigma^-$ hyperons.
These values of the bulk viscosity are in qualitative
agreement with those reported by Jones (\cite{jones01b}). 
If our model of bulk viscosity were more developed
and incorporated the contributions of process (\ref{Lambda}) 
involving $\Lambda$ hyperons
then the high-density regime 
would start to operate at somewhat earlier density,
at the $\Lambda$ hyperon threshold, $n_b = 0.310$ fm$^{-3}$.
The associated bulk viscosity is expected to be of nearly the same 
order of magnitude as produced by $\Sigma^-$ hyperons
(Jones \cite{jones01b}). Actually, in the presence
of hyperons, some contribution into the bulk viscosity
comes from modified and direct Urca processes
involving hyperons (e.g., Prakash et al.\ \cite{pplp92},
Yakovlev et al.\ \cite{ykgh01}). This contribution is not
shown in Fig.\ 1. It is expected to be smaller
than the contributions from nucleon modified and
direct Urca processes (\ref{Murca}) and (\ref{Durca})
displayed in the figure.

Figure 1 refers to one value of
temperature, $T=10^9$ K, and one value of the
vibration frequency, $\omega= 10^4$ s$^{-1}$. 
Nevertheless one can easily rescale
$\zeta$ to other $T$ and $\omega$ in non-superfluid matter 
in the high-frequency regime.
For the modified
Urca processes (M), direct Urca processes (D),
and hyperonic process ($\Sigma$) we obtain
the estimates:
\begin{eqnarray}
&&   \zeta_{\rm M} \sim {5 \times 10^{18}\, T_9^6 \over \omega_4^2}, 
   \quad \quad
   \zeta_{\rm D} \sim {5 \times 10^{24} \,T_9^4 \over \omega_4^2}, 
\nonumber \\
&&   \zeta_{\rm \Sigma} \sim {10^{30} \, T_9^2 \: \chi \over \omega_4^2}~~
   {\rm g \; cm^{-1} \; s^{-1}} .
\label{zeta-estimates}
\end{eqnarray}
The difference in magnitudes and temperature dependence
of $\zeta$ comes evidently from the difference of
corresponding reaction rates. It can be explained by different momentum
space restrictions (different numbers of particles, absence or presence
of neutrinos)
in these reactions (e.g., Yakovlev et al.\ \cite{ykgh01}).

\begin{figure}
\centering
\epsfysize=86mm
\epsffile[103 300 450 653 ]{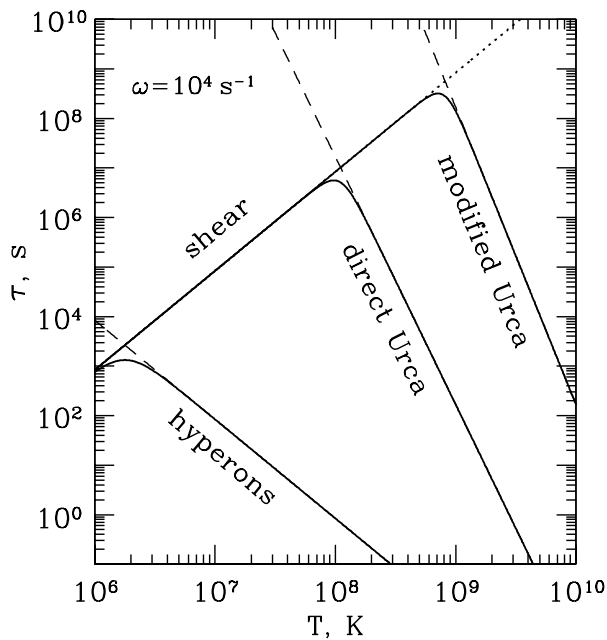}
\caption{
Schematic representation of temperature dependence
of viscous relaxation time scale $\tau$ in 
non-superfluid neutron
star cores with different compositions of matter at stellar vibration
frequency $\omega=10^4$ s$^{-1}$.
Three dashed lines show the relaxation 
due to high-frequency bulk viscosity associated either with
modified Urca processes, or with direct Urca processes, or
with hyperonic processes. 
The dotted line presents the relaxation due to shear viscosity.
Solid lines refer to the total (bulk+shear) viscous relaxation
for the three regimes.
}
\label{fig2}
\end{figure}

Now we can 
estimate viscous dissipation time scales $\tau$ of
neutron star vibrations.
A standard estimate based on hydrodynamic momentum-diffusion
equation yields
$\tau \sim \rho {\cal R}^2 /\zeta$,
where ${\cal R} \sim 10$ km is a radius of the neutron
star core, and  $\rho$ is a typical density.
For the leading processes of three types in
the high-frequency regime we have
\begin{equation}
   \tau_{\rm M} \sim {10\, \omega_4^2 \over T_9^6}~{\rm yrs}, \quad
   \tau_{\rm D} \sim {100\, \omega_4^2 \over T_9^4}~{\rm s}, \quad
   \tau_{\rm \Sigma} \sim {0.001\, \omega_4^2 \over T_9^2\, \chi}~{\rm s}.
\label{estimates}
\end{equation}
Schematic representation of the temperature dependence
of these time scales is shown in Fig.\ 2 by dashed lines.
Sharp difference of the dissipation time  scales comes
from different magnitudes of bulk viscosities in
various processes. In particular, the presence of
hyperons in the non-superfluid neutron star core results
in a very rapid viscous dissipation of stellar pulsations
(Langer \& Cameron \cite{lc69}, 
Jones \cite{jones71,jones01a,jones01b}).

Great difference of possible bulk-viscosity scales
is in striking contrast with the shear viscosity
limited by interparticle collisions. 
The shear viscosity $\eta$ should be rather insensitive to
composition of matter being of the same order
of magnitude as in $npe$ matter (Flowers \& Itoh \cite{fi79}),
i.e., $\eta \sim 10^{18} T_9^{-2}$ g cm$^{-1}$ s$^{-1}$.
It is independent of the pulsation frequency 
$\omega$. The damping time of stellar pulsations
via shear viscosity in a non-superfluid stellar core
is $\tau_{\rm shear} \sim 10 \, T_9^2$ yrs. 
It is shown in Fig.\ 2 by the dotted line.
This damping dominates at low $T$
while the damping by bulk viscosity dominates at higher $T$.
The total viscous damping time $\tau$
($\tau^{-1} \sim \tau_{\rm bulk}^{-1}+
\tau_{\rm shear}^{-1}$) is displayed in Fig.\ 2 by the solid
lines (for the three high-frequency bulk-viscosity damping regimes). 
One can easily show that
damping by bulk viscosity associated with modified Urca processes 
dominates at $T \ga 10^9\, \omega_4^{1/4}$ K.
For direct Urca processes it dominates at 
$T \ga 10^8 \, \omega_4^{1/3}$ K,
and for hyperonic processes at 
$T \ga 3 \times 10^6 \, \omega_4^{1/2}$ K. 

Finally, let us mention the validity of high-frequency
bulk viscosity regime. As follows from Eq.\ (\ref{hvis})
it is valid as long as $a \ga 1$, i.e., $\omega \ga \omega_c$,
where the threshold frequency $\omega_c \sim |\lambda| B/n_b$.
From Eq.\ (\ref{a}) for the hyperon bulk viscosity we
have $\omega_c^{\rm \Sigma} \sim 500 \, \chi \, T_9^2$ s$^{-1}$.
Using the results of Papers I and II we obtain
$\omega_c^{\rm M} \sim 5 \times 10^{-9} T_9^6$ s$^{-1}$ and
$\omega_c^{\rm D} \sim 5 \times 10^{-3} T_9^4$ s$^{-1}$ 
for modified and direct Urca processes. Therefore,
we always have the high-frequency
regime for modified and direct Urca processes at typical
temperatures $T \la 10^{10}$ K and 
pulsation frequencies $\omega \sim 10^4$ s$^{-1}$.
The same is true for hyperon bulk viscosity excluding
possibly the case of very hot plasma, $T \sim 10^{10}$ K.
Notice that in the low-frequency (static) limit
$\zeta \propto 1/|\lambda|$ and the temperature
dependence of the bulk viscosity 
is inverted with respect to the high-frequency case. 

\subsection{Superfluid reduction}

As discussed in detail in Papers I and II superfluidity
of nucleons can strongly suppress the bulk viscosity
produced by direct and modified Urca processes.
Now let us use the results of Sect.\ 3 and illustrate
superfluid suppression of hyperon bulk viscosity.

\begin{figure}
\centering
\epsfxsize=86mm
\epsffile[105 300 440 650]{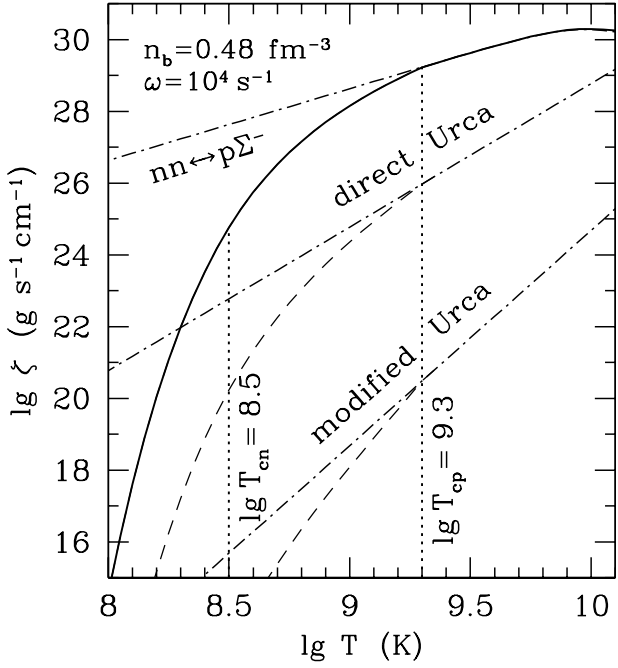}
\caption{
Temperature dependence of bulk viscosity at $n_b=0.48$ fm$^{-3}$
and $\omega=10^4$ s$^{-1}$ in the presence of proton superfluidity
with $\lg T_{cp}=9.3$ and neutron superfluidity with 
$\lg T_{cn}=8.5$. 
Dot-and-dashed lines (from up to down):
partial bulk viscosities due to hyperonic, direct Urca and
modified Urca processes, respectively, in non-superfluid matter.
Associated solid and dashed lines: 
the same bulk viscosities in
superfluid matter.
Vertical dotted lines show $\lg T_{cn}$ and $\lg T_{cp}$.
}
\label{fig3}
\end{figure}

Figure 3 shows this suppression 
at $n_b=0.48$ fm$^{-3}$ and
$\omega=10^4$ s$^{-1}$. We present partial
bulk viscosities produced by hyperonic processes, as well as by
direct and modified Urca processes.
The straight dot-and-dashed lines are the partial bulk
viscosities in non-superfluid matter.
The striking difference of these
bulk viscosities is discussed in Sect.\ 4.1. 
Solid and dashed lines show partial bulk viscosities
in matter with superfluid protons ($\lg T_{cp}{\rm [K]}=9.3$)
and neutrons ($\lg T_{cn}=8.5$). At $T \ga 10^{10}$ K
the high-frequency approximation for the
hyperon bulk viscosity is violated. One can see the tendency
of inversion of the temperature dependence
of $\zeta$ at $T \sim 10^{10}$ K 
associated with the transition to 
the low-frequency regime (Sect.\ 4.1).
At $T<T_{cp}$ superfluidity 
reduces all partial bulk viscosities. 
In the temperature range $T_{cn} < T < T_{cp}$,
where protons are superfluid alone,
all the three partial bulk viscosities are suppressed
in about the same manner. This is natural 
(e.g., Yakovlev et al.\ \cite{yls99}) since
the reactions responsible for the partial bulk viscosities
contain the same number of superfluid particles
(one proton). Indeed, there is one proton in hyperonic
reaction (\ref{Sigma}) and direct Urca reaction
(\ref{Durca}), as well as in the neutron
branch $N=n$ of
modified Urca reactions (\ref{Murca}).
At lower temperatures, $T < T_{cn}$, where neutrons
become superfluid in addition to protons, the suppression
is naturally stronger and becomes qualitatively different  
for different partial bulk viscosities since the leading
reactions involve
different numbers of neutrons. Evidently, the suppression
is stronger for larger number of superfluid particles 
(as well as
for higher critical temperatures $T_c$).

\begin{figure}
\centering
\epsfxsize=86mm
\epsffile[105 300 440 650]{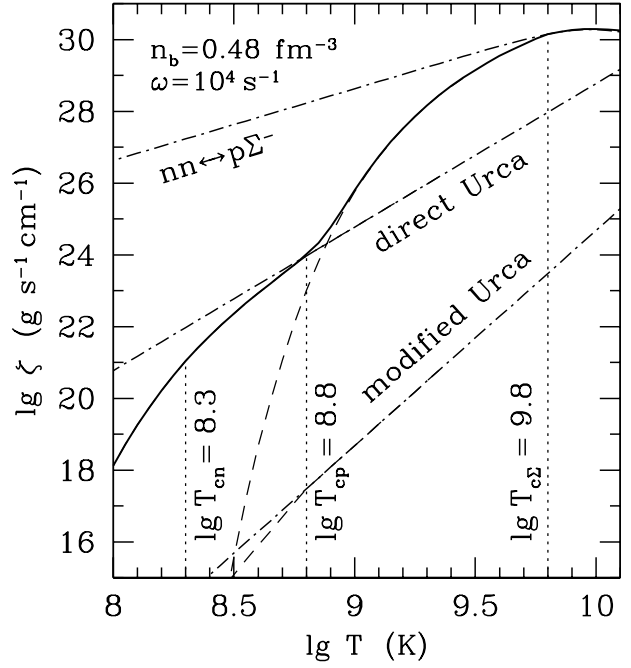}
\caption{
Same as in Fig.\ 2 but in the
presence of proton, neutron and $\Sigma^-$ superfluids
with $\lg T_{cp}=8.8$, 
$\lg T_{cn}=8.3$ and $\lg T_{c\Sigma}=9.8$. 
}
\label{fig4}
\end{figure}

Figure 4 exhibits the same temperature dependence
of the partial bulk viscosities, as Fig.\ 3, but 
in the presence of superfluidity of $n$, $p$, and $\Sigma^-$
($\lg T_{cp}=8.8$, $ \lg T_{c\Sigma} =9.8$, $\lg T_{cn}=8.3$).
One can see that superfluidity of $\Sigma^-$ hyperons strongly
reduces the partial 
bulk viscosity associated with hyperonic process.
As a result, at $T \la 10^9$ K the total bulk viscosity
is determined by direct Urca processes. In this regime
one should generally take into account the contribution from
direct Urca processes with hyperons (Sects.\ 1, 4.1). 
However, under the conditions displayed in Fig.\ 4 
this contribution can be neglected.  

Therefore, sufficiently strong superfluidity of baryons
may reduce the {\it high-frequency} 
bulk viscosity by many orders of magnitude.
This reduction will suppress very efficient viscous
damping of neutron star pulsations in the presence
of hyperons (Sect.\ 4.1). Accordingly, tuning
critical temperatures $T_c$ of different baryon species
one can obtain drastically different viscous relaxation times.

Note that relaxation in superfluid neutron star cores
may also be produced by a specific mechanism of {\it mutual
friction} (e.g., Alpar et al.\ \cite{als84},
Lindblom \& Mendell \cite{lm00}, and references therein). 
If the neutron star core 
is composed of $n$, $p$, $e$ (and possibly $\mu$), 
this mechanism requires superfluidity of 
neutrons and protons, as well as rapid stellar
rotation. The fact that the (conserved) particle 
currents are, in the case of a mixture of superfluids, 
not simply proportional to the superfluid velocities, 
implies non-dissipative 
{\it drag} (called also {\it entrainment}) 
of protons by neutrons.  
Dissipation (mutual friction)  is caused 
by the scattering of electrons (and muons) off the 
magnetic field induced by proton drag within the neutron 
vortices. The relaxation (damping) time associated 
with mutual friction, $\tau_{\rm mf}$, depends on the 
type of stellar pulsations and the
physical conditions within the superfluid neutron star core, 
in particular --- on the poorly known superfluid drag 
coefficient. Its typical value $\tau$ varies
from $\sim 1$~s to  $\sim 10^4~$s. 
One can expect that
similar mechanisms may operate in the superfluid hyperon
core of a rapidly rotating 
neutron star. If so these mechanisms will produce efficient
damping of stellar pulsations. Note, however,
that theoretical description of mutual friction
is complicated and contains many uncertainties.

\section{Conclusions}

We have proposed a simple solvable model (Sect.\ 2)
of the bulk viscosity of hyperonic matter in the neutron star
cores as produced by process (\ref{Sigma}) involving
$\Sigma^-$ hyperons. We have analyzed (Sect.\ 3) 
the hyperonic bulk viscosity 
in the presence of superfluids
of neutrons, protons, and $\Sigma^-$ hyperons.
We have presented illustrative examples (Sect.\ 4) 
of the bulk viscosity in non-superfluid and superfluid
neutron star cores using the equation of state of 
matter proposed by Glendenning (\cite{glendenning85}). 
In particular, we emphasized the existence of three
distinct layers of the core (outer, intermediate
and inner ones), where the bulk viscosity in non-superfluid matter
is very different (in agreement with the earlier
results of Jones \cite{jones71,jones01a,jones01b}).
This leads to very different viscous damping times
of neutron star vibrations for different neutron star
models (the presence or absence of hyperons; the presence or
absence of direct Urca process). 
If we used another
equation of state of hyperonic matter
the threshold densities $n_b$ of the appearance of
muons and hyperons, and the fractions of various particles
would be different but the principal conclusions would remain the same.
As seen from the results of this paper
and Papers I and II, the high-frequency bulk viscosities in all
three layers may be strongly reduced by superfluidity
of baryons. A strong superfluidity
may smear out large difference
of the bulk viscosities in different layers. In addition, it 
relaxes the conditions of the high-frequency regime.

Our consideration of the bulk viscosity in hyperonic matter
is approximate since we include
only one hyperonic process (\ref{Sigma})
(Sects.\ 1, 2.1, 2.4) characterized by one
phenomenological constant $\chi$. It would be
interesting to undertake microscopic calculations
of $\chi$. Analogous problem of quenching the axial-vector
constant of weak interaction in dense matter has been
considered recently by Carter \& Prakash (\cite{cp01}). 
It would also be important to determine
analogous constants for other hyperonic reactions
(e.g., for (\ref{Lambda})) 
in the dressed-particle approximation. This would allow one
to perform accurate microscopic calculations of the bulk
viscosity of hyperonic matter.

In Sect.\ 4.1 we have presented simple estimates
of typical bulk viscosities and associated damping time scales
of neutron star vibrations in different non-superfluid 
neutron star models. Let us stress that the 
actual decrements 
or increments of neutron star pulsations 
have to be determined numerically by solving an appropriate
eigenvalue problem taking into account various
dissipation and amplification mechanisms 
(e.g., bulk and shear viscosities;
mutual friction; gravitational radiation)
in all neutron star layers, proper boundary conditions,
etc. (e.g., Andersson \& Kokkotas \cite{ak00}).
In principle, the vibrational motion of various superfluids
may be partially decoupled. If so our
analysis of superfluid suppression of the bulk viscosity
must be modified (Sect.\ 3.2).
Nevertheless, the presented estimates 
and the theory of superfluid suppression show
that one can reach drastically different conclusions
on the dynamical evolution of neutron star vibrations
by adopting different equations of state in the neutron
star cores (with hyperons or without), different
superfluid models and neutron stars models
(different central densities, allowing or forbidding
the appearance of hyperons and/or operation of direct
Urca processes). We expect that the results of this paper
combined with the results of Papers I and II will be useful
one to analyze this wealth of theoretical scenarios.  

{\it Note added at the final submission}. After submitting
this paper to publication we became aware of the
paper by Lindblom \& Owen (\cite{lo01})
devoted to the effects of hyperon bulk viscosity
on neutron-star r-modes. The authors analyzed
the bulk viscosity taking into account
two hyperonic reactions, Eqs.\ (\ref{Sigma}) and (\ref{Lambda}),
and superfluidity of $\Sigma^-$ and $\Lambda$ hyperons
(but considering non-superfluid nucleons). 
Their treatment of the bulk viscosity in non-superfluid matter
is more general than in the present paper 
since they include the reaction 
(\ref{Lambda}). They treat the superfluid
effects using simplified reduction
factors which is less accurate (a comparison of analogous
exact and simplified reduction factors is discussed,
for instance, by Yakovlev et al.\ \cite{yls99}).
The principal conclusions on the main properties
of the bulk viscosity in hyperonic non-superfluid and
superfluid matter are the same.

\begin{acknowledgements}
Two of the authors (KPL and DGY) acknowledge
hospitality of N.\ Copernicus Astronomical
Center in Warsaw. The authors are grateful
to M.\ Gusakov who noticed a new form of the angular
integral, Eq.\ (\ref{Apl0}).  
This work was supported in part by the
RBRF (grant No. 99-02-18099),
and the KBN (grant No. 5 P03D 020 20). 
\end{acknowledgements}


\begin{thebibliography}{22}

\bibitem[1984]{als84}
Alpar, M. A., Langer, S. A., \& Sauls, J. A. 1984, ApJ 282, 533

\bibitem[2001]{ak00}
Andersson, N., \& Kokkotas, K. D. 2001, Int.\ J. Mod. Phys.\ D10, 381

\bibitem[2001]{bhy01}
Baiko, D. A., Haensel, P., \& Yakovlev, D. G. 2001, A\&A  374, 151

\bibitem[1998]{bb98}
Balberg, S., \& Barnea, N. 1998, Phys.\ Rev., C57, 409


\bibitem[2001]{cp01}
Carter, G. W., \& Prakash, M. 2001, Phys.\ Lett.\ B (submitted,
nucl-th/0106029)

\bibitem[1996]{dl96}
Dai, Z., \& Lu, T. 1996, Z.\ Phys.\ A335, 415

\bibitem[1968]{fw68}
Finzi, A., \& Wolf, R.A. 1968, ApJ 153, 835

\bibitem[1979]{fi79}
Flowers, E., \& Itoh, N. 1979, ApJ, 230, 847 

\bibitem[1985]{glendenning85}
Glendenning, N.K. 1985, ApJ 293, 470

\bibitem[1994]{goyaletal94}
Goyal, A., Gupta, V.K., Pragya, Anand, J.D. 1994, Z.\ Phys.\ A349, 93

\bibitem[1992]{hs92}
Haensel, P., \& Schaeffer, R. 1992, Phys.\ Rev., D45, 4708

\bibitem[2000]{hly00} 
Haensel, P., Levenfish, K. P., \& Yakovlev, D. G. 2000,
A\&A,  357, 1157 (Paper I)

\bibitem[2000]{hly01} 
Haensel, P., Levenfish, K. P., \& Yakovlev, D. G., 2001,
A\&A, 372, 130 (Paper II)

\bibitem[1971]{jones71}
Jones, P.B. 1971, Proc.\ Roy.\ Soc.\ Lond.\ 323, 111

\bibitem[2001a]{jones01a}
Jones, P.B. 2001a, Phys.\ Rev.\ Lett.\ 86, 1384

\bibitem[2001b]{jones01b}
Jones, P.B. 2001b, Phys.\ Rev.\ D64, 084003

\bibitem[1969]{lc69}
Langer, W. D., \& Cameron, A. G. W. 1969, Astrophys.\ Space Sci.  5,  213

\bibitem[1991]{lpph91}
Lattimer, J. M., Pethick, C. J., Prakash, M., \& Haensel, P.
1991, Phys.\ Rev.\ Lett., 66, 2701

\bibitem[2000]{lm00}
Lindblom, L., \& Mendell, G. 2000, Phys.\ Rev.\ D61, 104003

\bibitem[2001]{lo01}
Linblom, L., \& Owen, B.~J. 2001, Phys.\ Rev.\ D (submitted,
astro-ph/0110558)

\bibitem[2001]{ls01}
Lombardo, U., \& Schulze, H.-J. 2001,
in Physics of Neutron Star Interiors,
eds.\ D.\ Blaschke, N.\ Glendenning, A.\ Sedrakian
(Springer, Berlin) p. 31.

\bibitem[1992]{madsen92}
Madsen, J. 1992, Phys.\ Rev.\ D46, 3290

\bibitem[1992]{pplp92}
Prakash, M., Prakash, M., Lattimer, J.M.,  \& Pethick C.J.
1992, ApJ 390, L77

\bibitem[1989a]{sawyer89}
Sawyer, R.F. 1989a, Phys.\ Rev.\ D39, 3804

\bibitem[1989b]{sawyer89b}
Sawyer, R.F. 1989b, Phys.\ Lett.\ B233, 412 

\bibitem[1997]{sw97}
Savage, M.J., \& Walden, J., 1997, Phys.\ Rev.\ D55, 5376

\bibitem[1983]{st83} Shapiro, S. L., \& Teukolsky, S. A. 1983,
Black Holes, White Dwarfs and Neutron Stars, Wiley-Interscience,
New-York

\bibitem[1984]{wl84}
Wang, Q.~D., \& Lu, T. 1984, Phys.\ Lett.\ B148, 211

\bibitem[1999]{yls99}
Yakovlev, D. G., Levenfish, K. P., \& Shibanov Yu. A. 1999,
Physics-Uspekhi, 42, 737 [arXiv: astro-ph/9906456]

\bibitem[2001]{ykgh01}
Yakovlev, D. G., Kaminker, A. D., Gnedin, O. Y., \& Haensel, P. 2001,
Phys.\ Rep.\ 354, 1  

\end{thebibliography}
\end{document}